# Modeling of Resource Allocation Mechanisms in Distributed Computing Systems using Petri Nets and Stochastic Activity Networks (SAN): a Review and Reo-based Suggestion

Mohammad Reza Besharati, PhD Candidate, Sharif University of Technology, Tehran, Iran, besharati@ce.sharif.edu,
Ali Sepehri Khameneh, Graduated Master, Sharif University of Technology, Tehran, Iran

1. **Abstract:**

   
   Resource allocation is crucial in the distributed systems. It is a key step in designing the mechanisms of systems for determining the resource allocation mechanism, it is important for obtaining the desired efficiency in the system, plus it is vital for predicting and preventing Deadlocks. Various models of Petri Net (Stochastic PN, Colored PN, Generalized PN, etc.) are used for modeling, simulation, execution, and solving the problems of resource allocation. SAN models are used for modeling the problems pertinent to resource allocation. First, we shall address the basic concepts pertinent to these models and the resource allocation problem (introduction chapter), then, some applications of the Petri Net and SAN models in the distributed computational systems or systems based on them shall be studied. Finally, the issues and findings will be concluded.

2. **Abstract of Findings**

   Resource allocation in the distributed systems is a crucial and principle issue. Modeling using Petri-Net and SAN networks (and some others such as formal automata models and etc.) is among the modeling methods of the resource allocation systems. Using this modeling method these systems can be analyzed and simulated and therefore, assess and compare the different design solutions for these systems.

   A research concern in this regard is finding Petri-Net classes proportionate to the classes of the resource allocation systems. It must be emphasized that most of the systems around the world, at least in the specific level of abstraction, can be reduced to a system of resource allocation to analyze and assess their designs through Petri-Net modeling. In most of these systems, Deadlock is the main problem and concern. Deadlock concerns, such as predicting or preventing its occurrence, can be investigated through analysis of the Petri-Net model or simulation based on that model.

   Sometimes, the problem of resource allocation exists in a platform of limitation or requirements. Meaning that you cannot do as you please. For instance, we might have a resource management system that besides resource allocation, it is required to manage the problem of scheduling jobs and reliability as well. In such systems, the problem of resource allocation must be resolved together and in interaction with other problems so that it can lead to limitations in the resource allocation problem. The more the limitations the more the complexity of the resource allocation problem, and the significance of simulation or analysis increases for management of this added complexity. Petri-Net enabled us to analyze the model and simulate based on the model.

One of the concerns in the resource allocation methods is that these methods are Qos-aware. Using Petri-Net and SAN modeling provides us with the opportunity to decide whether the method used for resource allocation is as good as Qos-aware or not through modeling and simulation.

Another problem is the amount of Robustness of the resource allocation method. Petri-Net modeling can be used again and simulating the amount of robustness of the resource allocation method can be determined.

In some distribution systems, controlling the executive operations of systems for preventing a fault is another concern. Here, instead of supervising and controlling the whole system, we can use its Petri-Net model and carry out the required supervision on the system, either by measurement or simulation.

Synthesizing the protocols in the distributed systems is one of the other fields of application of the Petri-Net modeling.

## 3. Introduction:

### 3.1. Petri-Net Modeling of Systems & Its Applications

#### 3.1.1. Petri Nets

In this chapter, we will introduce the Petri-Net model based on reference 1. The content of this chapter is provided from reference 1:

"The place/transition network in short Petri Net is defined as N=<P,T,W>. In this definition W as expressed as W: $(P \times T) \cup (T \times P) \to IN$. In this definition, P and T must be nonempty, limited, and separated from each other. P signified places and T signified Transactions.

In Petri Net the set of input nodes of node x is called preset, and the set of the output nodes of x. are called poset nodes, Preset is shown by •x and defined as •x = {y ∈ P ∪ T | W (y, x)≠0} and poset is shown by x• and defined as • x = {y ∈ P ∪ T | W (x, y)≠0}. Preset of X set that is X ⊆ P ∪ T is indicated by •X and defined as •X = {y| y ∈ • x, x ∈ X}. Poset of the X set is indicated by X• and defined as •X = {y| y ∈ x•, x ∈ X}. The ordinary Petri Net is normally shown by a single-weighed edge, in other words, W was defined as W: $(P \times T) \cup (T \times P) \to \{0,1\}$, in other words, Petri Net is called a public network. The state machine is a type of simple Petri Net, in which the size of preset and poset equals 1 at any state. A finite state machine is a simple Petri Net as well, in which the size of preset and poset is less than 1 at any state.

If p∈P node has a path it means that p∈ p••. A Petri Net aside from loop is a network that its node has no self-loop and is defined as N=<P,T,C>. In this definition C is called incidence matrix, which is defined as C[p,t] = W(p,t) – W(t,p).

A p-flow is a vector, which is a member of a power set of Z set and this set opposes zero and in case of multiplying this vector by the incidence matrix, the answer will equal 0.

A set of places is called D ⊆P‹ siphon if any p that is a member of the D presets (p ⊆•D), and be the member of the D posets.

m is called a marked Petri Net, which is the subset of the power set of N (natural numbers). A limited number is assigned to every box of m that signifies a p. ||M|| signifies a set of ps with a non-zero value in m vector. The marked Petri Net is defined as a pair <N,$m_0$>. $m_0$ is called for the initial marked Petri Net".[1]

### 3.1.2. Applications of Petri Net Modeling

Petri Net modeling of a system enables the users to predict aspects of system behavior through its Petri Net behavior. For instance, the Siphon and Trap structures defined in Petri Net that though them the occurrence of Deadlock and survival of the process can be predicted during some states (Reference 2). Even though making the decision that whether a Petri Net contains Siphon is itself an NP-Complete problem (Reference 3).

Furthermore, considering that Petri Net inherently supports the property of composability, it can provide a platform for managing the complexity of the workflows or resolving any problem containing the concept of process. Meaning that at first, complexity management and problem solving will be carried out on the subprocess and microprocesses, then, the higher processes will be obtained by composing the subprocesses. This property of Petri Net can be used to resolve the problem of extraction of the protocol entities for nodes of a Grid network.

### 3.2. SAN Modeling of Systems & Its Applications

To increase the expression capability, extensions from Stochastic Petri Net are provided that finally lead to the appearance of the Stochastic Activity Networks also known as SAN models. The structural components of SAN models include entrance gate, exit gate, selection symbol (Case for modeling the probability-based selections), the symbol of temporal activity, and the symbol of instantaneous activity. (Reference 4)

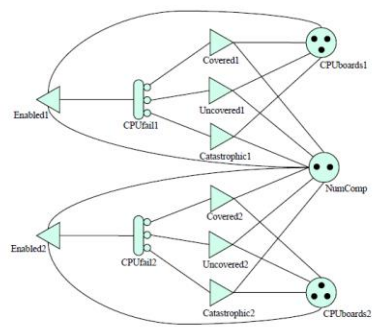

**Figure 1- An Instance of a SAN Model (Reference 4).**

---

[1] The section inside quotation mark is a translation of some parts of reference 1.

## 3.3. An Instance of Modeling of the System

The content of this chapter is provided by translating reference 1. This example includes modeling the Dining Philosophers Problem using Petri Net.

"In this instance, the famous example of Dining Philosophers is included. In this instance, 5 philosophers spend all of their time thinking and eating. These philosophers shared a round table and a dining room. At the middle of the table, there is a great bowl of spaghetti that is filled constantly. And there are five forks on the table as well. When a philosopher feels hungry, they go to the dining room and sit on their chair and take the forks on their right and left. Then, measures the temperature of the spaghetti, if the spaghetti is cold, the philosopher leaves the forks on the tables and puts the spaghetti in the microwave. After the spaghetti is heated to a sufficient amount, the philosopher returns to the dining room, sits on their specific chair, takes the left fork, and puts the spaghetti on the table. However, since the spaghetti is intertwined, thus, the philosopher must use the fork on their right too to take some spaghetti. In case the philosopher takes the fork before the spaghetti is cold, they can start eating. When finished eating, places the forks on the table and leaves the room.

Based on the old nomenclatures, philosophers are the periodic threads and the forks and the bowl of spaghetti are the reusable resources shared among the five philosophers. From a software point of view, each philosopher is a thread or a thread that is executed simultaneously in the system.

The algorithm in Figure 1 proposes a code for resolving this problem. The method of acquiring and releasing the resources is used to solve this problem. For acquiring (the wait) function and for releasing (the signal) function are used. Both of them are popularized for acquiring and releasing resources. The (trywait) function is a function similar to (wait) without block. The respective resource is sent to this function and in case this resource is accessible, the resource is acquired and returns the TRUE value, otherwise, this function will return the FALSE value. For simplicity, it is assumed that in the condition in which two or resources are assessed, this state is carried out automatically. In Figure 1 the network pertinent to the algorithm is shown as i=1, in this network the details pertinent to resource allocation are deleted. Figure 2 shows the same network by composing the details of resource sharing. Bear in mind that if we eliminate the dashed line from Figure 2, we will have fine independent connected machines that are isolated with six places. Each machine shows the state of a control flow for each philosopher. Every state machine comprises seven states. Token shows the concurrency process in the state machine that shares the control of the flow. At the initial state, all philosophers are thinking (outside the dining room). In each state machine, the token in an idle state. In general, an idle state is a mechanism, through which the number of threads in the process decreases. Here, at every moment only one i philosopher can enter the dining room. Six isolated states are called resource states. A resource place presents a type of resource and the tokens inside it show the accessible samples of that resource. At this state, every resource place is marked solely. Therefore, at the initial state, an i fork, which is the member of {1, 2, 3, 4, 5} exists together with a bowl of spaghetti.

```
var
    fork: array [1..5] of semaphores; // shared resources
    bowl: semaphore;                  // shared resource
begin
    do while (1)
        THINK;
        Enter the room;
(T1)    wait(fork[i]);                       (T6)
        do while (not(trywait(bowl, fork[i mod 5 +1]))
                  or the spaghetti is cold)
(T2)        if (trywait(bowl)
                and the spaghetti is cold) then
(T3)            signal(fork[i]);
                Go to the microwave;
                Heat up spaghetti;
                Go back to table;
(T4)            wait(fork[i]);
(T5)            signal(bowl);
            end if;
        loop;
        Serve spaghetti;
(T7)    signal(bowl);
        EAT;
(T8)    signal(fork[i], fork[i mod 5 +1]);
        Leave the room;
    loop;
```

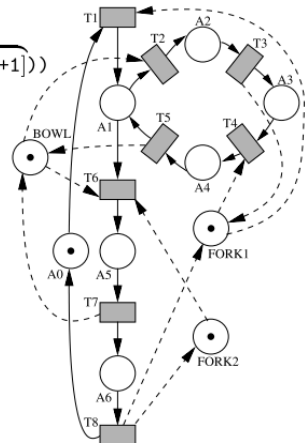

Fig. 1. Philosopher 1

**Figure 2- Algorithm for Dining Philosophers Problem (Reference 1).**

Finally, the dashed arrows show the acquiring and releasing of the resources by the threads when they are changing the state. When the transaction starts working, the accessible resources change. Bear in mind that, in a state in which an isolated token is moving inside the state machine, you should disregard other unchanged states until its return to the idle state. Thus, the resources are used protectively.

Some of the capabilities of RAS models will be discussed that the concurrent program should contain. A state machine without a loop inside is not considered a universal machine in the field of FSM. The $S^3PR$ and $S^4PR$ classes upgrade the capability of the state machine every time that the cycles pass an idle state. Even though, this item can lead to limitations even in very simple systems. In their theory, Bohm and Jacoini stated that every non-structured program can be transformed into a structured program by reconstruction. Consequently, we can restrict the models of the process to states in which decision and loop are used using every branch is not authorized. Another difference between FSM and a software system is that this system does not require physical resources and the resources can exist logically. This is a great concept in resource allocation. A resource is an object that is commonly used by the threads and this usage must be exclusive. When the number of resources is finite, the processes compete for using the resources and in case of access to a resource maintain it until the end. Since resources are allocated in this system to enable the thread to maintain the resource by its side until the end of its work, therefore there might be a deadlock. Otherwise, the resources can be protected using binary mutex, lock, and semaphore to protect single resources and the counting semaphore is used to protect multiple sample resources. These resources can be diverse, however, the required concurrency will be the same. If we can use these semaphores in non-exclusive systems they can be defined properly. For instance, the counting semaphore can be used to control the number of connections for a database. In this state, the threads wait on the semaphore until another thread decides to close its connection with the database. The semaphore has the intermediary role for facilitating the coordination, besides, the can lead to blocking as well. In this article (the reference article) at first, a solution is mentioned for solving the deadlock problem at resource allocation state. Then, this solution will be mapped to the message transfer state in order to use this solution for solving that problem as well.

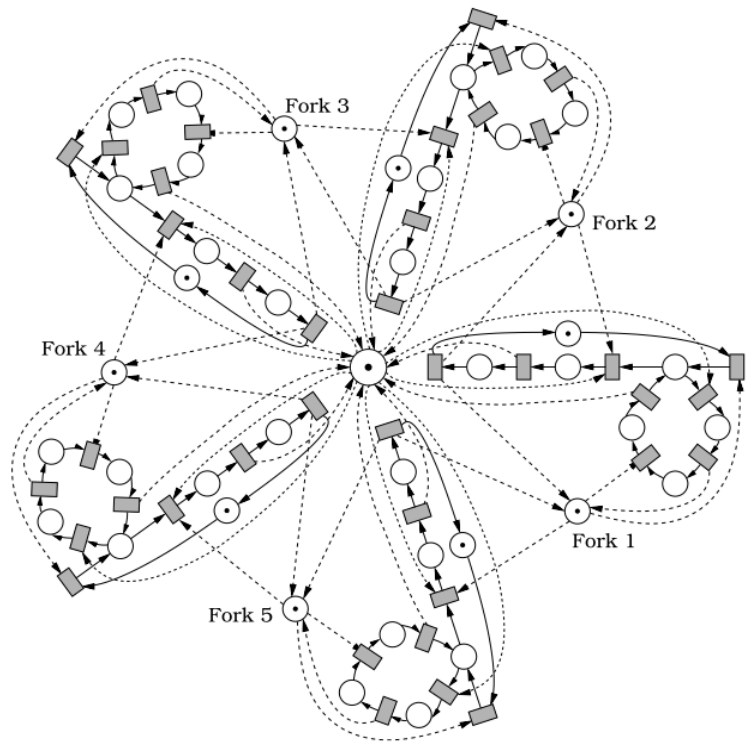

**Figure 3- Petri Net, Thinking Philosophers (Reference 1).**

Due to its diversity, a semaphore can be used to determine the process of resource allocation by the threads. For example, the XSI semaphores can be used for multiple exclusive (Oligopoly) wait. An XSI can be reduced automatically in several units. The XSI and POSIX semaphores both possess non-blocking waits.

In this field, the threads can not only use the resources, by creating them. The thread will be destroyed before the completion of the resource is created. For instance, a thread can create a shared variable and grant it to another thread. Thus, this resource allocation is not first-acquire-later-release, however, it can be solved through other methods. All resources are used protected by the thread (either by creation or destruction of the resources or by wait and acquiring the resources).

Thus, for successful modeling of the RAS system, the Petri Net must meet the following requirements:

1. The flow of the state of a thread must be modeled using a state machine that supports decision-making and internal cycles.
2. In this system, we can have several types of resources and several samples of each resource.
3. The state machine can have several states (supports concurrency).
4. Threads use the resources protectively.

5. The resource acquiring and releasing edges can have unusual weight (semaphores can be increased or decreased automatically and obtain a value of more than one).
6. The process of resource acquiring and releasing must be possible automatically.
7. The threads can make decisions based on the resource request (due to the request of a resource that does not lead to blocking).
8. The threads can rent their resources. As one of the impacts of this action, there must be resource-dependent threads that must be created and rented by another thread".[2]

## 3.4. Resource Allocation Problem & Resource Allocation Systems

The content of this chapter is provided by translating reference 25.

"Among the most repetitive patterns in the wide range of the engineering fields, the competition for common resources between the concurrent threads is considerably important. The perspective of the discrete-event system theory as a framework in which solutions are provided for resource allocation problems (RAP) appears to be suitable and strong. Systems like this are called resource allocation systems (RAS).

Resource allocation systems are investigated for two distinguished entities of the threads and resources. The objective of RAP is to successfully meet the resource requests by the threads so that no thread will ever face deadlock. When we say a set of threads reached a deadlock, it means that they wait for resources for an indefinite time that its resources are acquired by other threads in the same set.

Sequential RAS is a type of RAS that manages the resources that can be reused sequentially. Therefore, the execution of a thread can increase or decrease the number of free resources. The resources are used protectively in these systems.

Even though other concurrency models are considered as well, Petri Nets obtained a pioneering role in the official models used for dealing with RAP. One of the strengths of this strategy is the plane map between the main entities of RAS and the primary elements of Petri Nets. A type of resource can be used by a model place, its samples will be modeled with the tokens. Besides, the consecutive threads are modeled by tokens operate between the states. The direction of edges from the resource places to transfers (from transfers to resource places) is an indication of acquiring (releasing) some resources by the thread. Besides, their own advantages these Petri Nets provide a natural official framework for the analysis of RAS.

This truth in the domain of the flexible manufacturing system (FMS) that Petri Net models achieved great success for RAS since the work of Ezpeleta et al., was unveiled is known better. This is stabilized on two strong columns: 1. The definition of a rich syntax from a physical point of view; 2. using the precise scientific results enables us to distinguish the deadlocks from the model structure and have a well-defined methodology for their automatic modification in the system.

Nowadays, there are many Petri Net models for modeling RAS in the field of FMS that normally overcome the syntactic limitations of the S3PR class. The S4PR networks popularize the previous issue besides, enable the allocation of the concurrent multiple resources by a thread. The S*PR networks expand the expressive power of the threads as much as the state machines. Other classes such as NS-RAP, ERCN merged nets, PNR nets, expand the capabilities of S3PR/S4PR models more than consecutive RAS through separation or merging operations.

---

[2] This section was written based on the translations of the reference 1.

One of the analysis and control techniques in the research are on the basis of a structural element that specified deadlocks in most of the models: Which is known as bad siphon. A bad siphon does not support a p-semiflow. If the bad siphons are empty, the transfers of their outputs ends since the resource places can no longer accept the tokens, thus, they show deadly acceptance. The control techniques emphasize supervision places that restrict leakage of the tokens from the bad siphons.

Even though there is an obvious similarity between RAP in FMS and in the concurrent or parallel software, the previous efforts for including the already known RAS techniques to software engineering field have been limited and unsuccessful to the best of our knowledge.

One of the crucial concepts pertinent to resource allocation is the deadlock concept. The important question in the Petri Net theory is that "Is network in deadlock?" A Petri net has reached deadlock if there is no active transition. Here, two simple network models will be presented that help introducing the concept of deadlock.

The following model is used for introducing the concept of deadlock. In Figure 1, a simple Petri Net with the objective of resource allocation is demonstrated that comprises merely two single samples and two threads, both of which require the resources before completion. R place shows the single resource and each token shows a sample of the resource. The rest of the places manifest the control thread of two threads. The $t_1$ and $t_2$ transitions, show the events of a and b threads, respectively, that request a sample from the resource. The $t_3$ and $t_4$ transitions show the secondary request of each of them. Finally, the $t_5$ and $t_6$ transitions, model the release of both resource samples. If the order of $t_1$, then, $t_2$ occurs, the network will reach a deadlock. At this point, there is no active transition. Even though the resource allocation diagrams are beneficial, the Petri Net model adds the dynamic aspects explicitly, meaning that the possible order of occurrence that can lead to deadlock. The order of events that lead to deadlock can be tested.

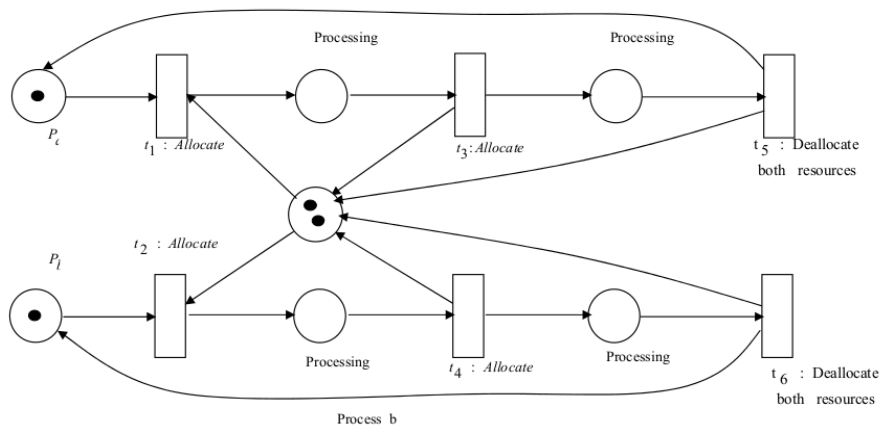

**Figure 4 - Resource Allocation Petri Net Model (Reference 25)**

When the first model is understood, the concept of deadlock is clear, the issue of preventing deadlock and the respective costs can be discussed. To show the preventive method in which the threads should allocate their resources before the commencement of the execution, the Petri Net of Figure 2 can be used. If $t_1$ is executed, $t_2$ and $t_4$ cannot be executed until the $t_3$ is executed as well. This fact is modeled that the thread a acquires all resources and b cannot proceed. After execution of $t_3$, both $t_1$ and $t_2$ become active.

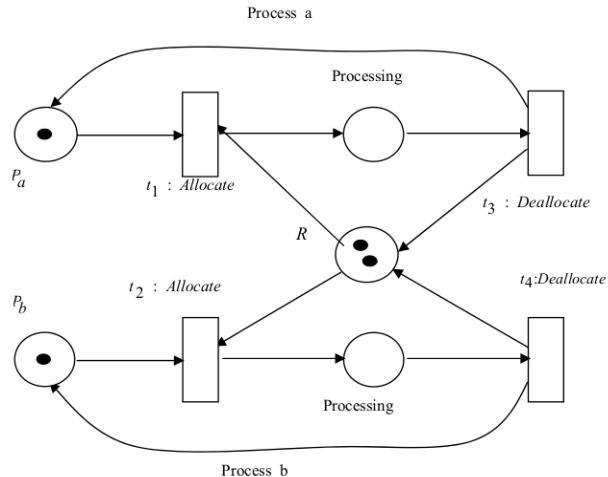

**Figure 5- Resource Allocation Petri Net 2 (Reference 25)**

Preventing deadlock and detection are other two great subthemes. When talking about preventing deadlock, the Banker's Algorithm is proposed. When there is a request by thread $p_i$, the Banker's Algorithm examines the security of the request. Another similar algorithm is used for the detection of deadlock. The algorithm of deadlock detection is executed consecutively. The output of the detection algorithm is either a set of threads that identify the manner of completion of the threads, therefore, it identifies that the systems have not reached a deadlock, or it identifies that the system has reached a deadlock.

A Petri Net was proposed for a deadlock detection algorithm for a system of threads with multiple samples of each resource (in the reference article of this discussion). In the following discussion, X is used as a non-ordinary index variable in order to use letters instead of non-negative integers. In case we have m threads and n resources, the algorithm uses an allocation matrix $n \times m$ called Allocation, in which $Allocation[i, X] = k$ means that k samples are allocated to i thread from the X resource. Moreover, a request $n \times m$ matrix called Request is considered, in which $Request[i, K] = k$ means that thread i has requested k samples from the X resource. There is an m-numbered vector named Available that if $Available[X] = k$, then it means that there is currently k samples available from the X resource. The model of the public network is described as such. There is a thread place and a pair of transitions for each thread of the system (all of them are explained). Furthermore, for each type of resource, there is a resource place. Assume that $p_i$ and $p_x$ show the place of thread and resource, respectively. The set of F edges and W labeling function include the following:

- $(P_i, t_i)$ Edges, each of them weighing one, for each i thread.
- Edges from $t_i$ transition to $p_x$ resource places with the weight label of k, in case of $Allocation[i, X] = k$ and $k > 0$.

- Edges from the resource places of $P_x$ to $t_i$ transition with the weight label of k, in case of $Request[i,X]=k$ and $k>0$.

In the algorithm, an n-numbered Boolean vector named Finish is used in order to maintain which threads are examined and can achieve their request. If the i thread is examined by the algorithm and it can be executed, then, $Finish[i]=true$. At first, for every i that $1\leqslant i\leqslant n$, $Finish[i]=false$.

The symbols of the place are proportionate to Finish. The thread $P_i$ is marked with a token, if $Finish[i]=0$, otherwise, it will be marked with zero tokens. After a $t_i$ transition is turned on, a token will be eliminated from the place of $P_i$ thread. Which means that we equal $Finish[i]$ to truth. The concept of turning on a transition is explained hereinbelow. The $P_x$ resource places are marked with k tokens, if $Available[X]=k$. Besides the data structure discussed earlier, the algorithm uses an m-numbered vector called Work, which is changed during the execution of the algorithm. First Work is considered to equal Available. At every point of algorithm execution, the value of $Work[X]$ equal the number of tokens in $P_x$. When the X resource, $Request[i,X]\leqslant Work[X]$ and $Finish[i]=false$, the transition equal to $t_i$ will be activated. It equals a situation in which the number i requests of the thread can be accepted. Bear in mind that more than one request can be accepted. One of them is accepted indefinitely. In case the i thread is selected, the Work vector is updated by vector extension of the number i line of Allocation. Which means turning on the $t_i$ transition.

Assume that there are five $\{p_0, p_1,..p_4\}$ threads and three types of $\{A,B,C\}$ resources. Resources A, B, and C contain 7, 2, and 6 samples, respectively. Assume that the matrix that shows the allocation, request, and entity status are demonstrated at T moment in Table 1.

Table 1: Threads, various types of resources, and requests

| | Allocation | | | Request | | | Available | | |
|---|---|---|---|---|---|---|---|---|---|
| | A | B | C | A | B | C | A | B | C |
| $p_0$ | 0 | 1 | 0 | 0 | 0 | 0 | 0 | 0 | 0 |
| $p_1$ | 2 | 0 | 0 | 2 | 0 | 2 | | | |
| $p_2$ | 3 | 0 | 3 | 0 | 0 | 0 | | | |
| $p_3$ | 2 | 1 | 1 | 1 | 0 | 0 | | | |
| $p_4$ | 0 | 0 | 2 | 0 | 0 | 2 | | | |

The model of the network of Figure 3 enables illustration of the deadlock detection through the graphical algorithm. In this model, the indefinite activated transitions will be on until there is no active transition left.

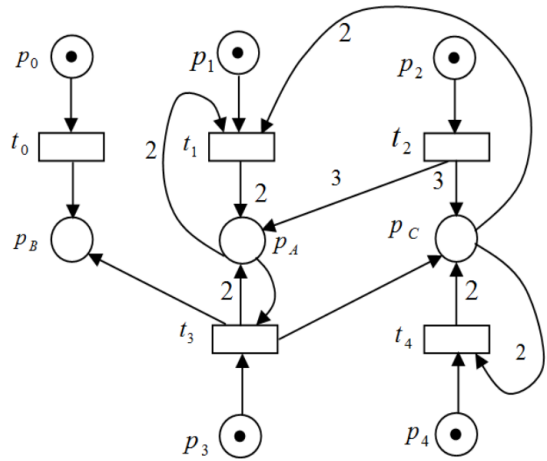

**Figure 6 - Deadlock Detection Model (Reference 25)**

In this model, it does not mean that the system has reached the deadlock, the deadlock is modeled, it will be detected in case a token has remained in one of the primary specified thread places and the is no active transition. If all of these thread places are unmarked, in that case, the order of turning on equals an order of thread completion. For the above example, $t_0$, $t_2$, $t_3$, $t_1$, and $t_4$ is a possible turning on order. It shows that the above system is not in a deadlock state. Another example of the possible orders are as follows: $t_2$, $t_4$, $t_0$, $t_3$, and $t_1$.

The proposed Petri Net model in the following chapter is another graphical technique for detection of deadlocks and as demonstrated in this article (the reference article) it can be used for modeling other issues in the operating system".[3]

## 4. Some Applications of Petri Net and SAN Modeling in the Computational Distributed Systems for Resource Allocation

### 4.1. Applications in Grid Systems & Environments

One of the applications of the Grids is carrying out scientific virtual tests and computations. The complexity of the scientific computations and tests can be modeled in form of workflow and Petri Net is one of the workflow modeling tools. Thus, Petri Nets are crucial for mass scientific computations on the Grids. One of the problems that emerge in these computations is the limitation of the resources that can be modeled by the Petri Nets. Therefore, the Petri Nets are used for modeling in two levels (the logical level pertinent to a model that computation is based on and a level that shows the limitations of the resources of the computing system during computation). Therefore, the Petri Nets' hierarchical modeling and multiple layers are important as well. This modeling method can be important with respect to other aspects.

---
[3] This subsection was written based on the translations of the reference 25.

Another problem is that deadlocks can occur at both levels mentioned earlier. Meaning that the logic of the workflow is computed and the model that the computation is based on will have deadlock (the complex processes and workflow proportionate to them, which is simple). Therefore, assessment of the process of computation, a model to carry out the computation based on it, and ensuring that this model will have no deadlocks is quite important. Aside from that, the computational resources of Grid must not be dependant so that during the execution of the computation, there shall be no deadlocks on the Grid itself. Thus, as mentioned in both of the earlier levels, the occurrence of deadlock is possible and it is not pleasant. Since Petri Nets possess proper formalism and defined algorithms for the prediction of the deadlocks on their structure, they are important as well. Therefore, an article suggested using Petri Nets for the description of a workflow on a Grid and enjoy these benefits. The rival models, such as DAGs or descriptor Scripts, have either less expressive ability or they are complicated and they do not have the simplicity of Petri Nets, considering goals such as deadlock occurrence.

In the field of virtual organizations, one of the distributed computing systems is the grid workflow-based systems. A grid workflow is the result of the combination of separate activities that are executed on different processing resources, job dependent [4], and distributed in the format of a virtual organization. Taking into account that the scheduling problem for such systems is NP-Hard, the heuristic search algorithms executed in Time Petri Net models can be used for solving this problem (Reference 6).

In the Grid environments, the resource management system (RMS) is responsible for scheduling and executing the jobs. RMS breaks the input jobs into several sub-jobs, distributes them among the resources in the Grid environment, receives the results from the environment, and delivers as output. Scheduling by RMS and execution of jobs by the resources can be modeled using Colored Petri Net models. Analysis of dependability of the Grid environment can be carried out by analyzing these models (Reference 7).

For modeling the job scheduling and analysis of the efficiency of the Grids the Stochastic Petri Net can be used. Besides, for an Availability environment, the Grid of the resource management system was quite crucial. SAN models are employed to analyze the Availability of these systems (Reference 8) and (Reference 9).

In another article analyzing the Availability of a Grid environment at a point in time on the basis of the SAN model was investigated and it was stated that the mass of the resources available in a Grid system can lead to loss of Availability at a point in time. It was also mentioned that the results of this modeling can be used to improve the Grid policies and select the best of them. It was also expressed that a better resource allocation will be carried out. After analyzing the Availability of a Grid environment from the SAN model, better allocation of the resources can be achieved (Reference 10).

The Petri Nets are used for simulation of job allocation in the distributed computational systems as well. In this resource, a Generalized Stochastic Petri Net is used for modeling the dependencies in the graph of the jobs. A map is defined and established between each graph and job and Generalized Stochastic Petri Net. The problem of resource allocation causes dependence among the jobs (threads) that can be modeled by Generalized Stochastic Petri Net. The benefit of Generalized Stochastic Petri Net supports the modeling of the schooled transitions, besides, coding of the not scheduled transitions (Reference 11).

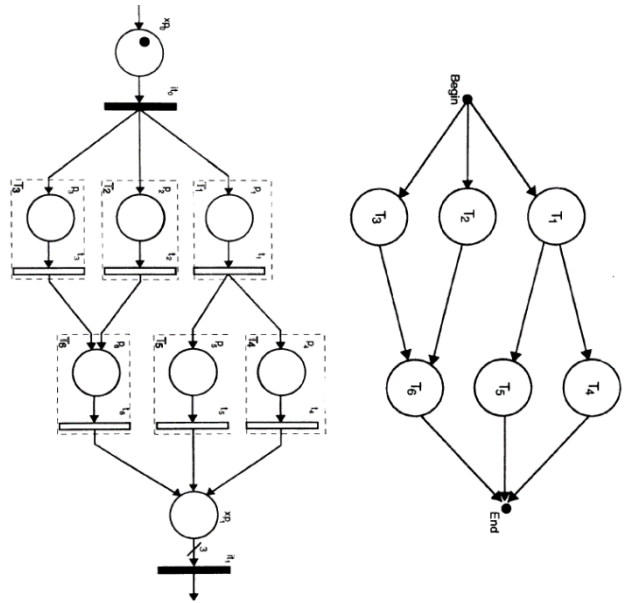

**Figure 7 - A Job Dependency Graph (Right) and a similar Petri Net (Left) (Reference 11)**

In another article, Petri Net was used for modeling the workflow, carrying out schedules based on that Grid system. These samples are completely applicable and operational samples of the application of Petri Net for modeling and resource allocation in a Grid system (Reference 12).

### 4.2. Application of WorkFlow Systems

As observed in the chapter on applications in Grid systems & environments, Petri Nets are used for complexity management of processes in the WorkFlow systems. To prevent the repetition of the subjects, please refer to the respective section. Besides the discussed issues in that section, there are access limits for various resources or users due to organizational considerations or access security, which adds another layer of complexity to the workflow (Reference 13). The complexity of the workflow can be managed by Petri Nets.

### 4.3. Application of Multi-Agent Systems

Sharing the resources among the agents in an agent-oriented system is an important research subject. In a specific approach, sharing resources among the agents is managed through specific components called the coordinator. During the design of an agent-orients system, there are a variety of choices for designing and configuration of the coordinators. To select the most suitable option for coordinators, simulation based on Colored Petri Net models is used. Therefore, the best mechanism for management of the shared resources among the agents in an agent-orients system (a distributed system as well) can be selected and adjusted through a simulation approach (Reference 14).

Another approach of the agent-oriented architecture is used to resolve the problem of resource allocation. A hierarchy of the agents is defined that is responsible for the long-term, medium-term, and instant resource allocation job. The logic and work process of each agent is described and modeled through the Petri Net model. The combination of the operation of agents equals a combination of the Petri Net models.

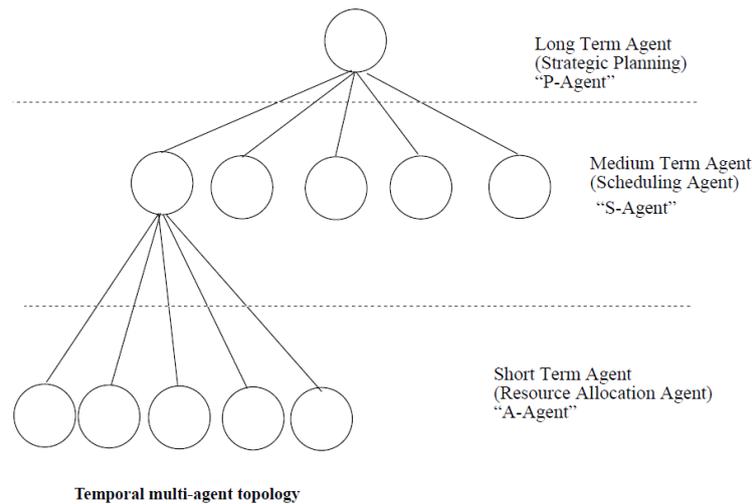

**Figure 8 - (Reference 15)**

In this approach, the problem of resource allocation by agents is resolved. In fact, the executive performance of the agents results in resolving the resource allocation problem. However, the executive performance of the agents are in fact the execution of Petri Net models that describe and model each agent. Thus, in this approach through the execution of the Petri Net models the problem of resource allocation is resolved. It must be taken into account that these Petri Nets possess the logic for problem-solving and its resolving policies not describing the system for carrying out the resource allocation (Reference 15).

The problem of fault tolerance in a distributed agent-oriented system is considered as a resource allocation problem: We just want a limited amount of replication for agents. For which agents we consider replication for the agent: The resources of replication for the agent are allocated to which agents? An approach for resolving this problem is Plan-Based Resource Allocation. In this approach, the design and plan can be cab used as other Petri Net (Recursive Petri Net). The difference between these models and ordinary Petri Net models is not the singular actions and it can be an abstraction of composed actions. The Petri Nets are used for modeling the logic for solving the problem of resource allocation and not modeling the system that we are trying to carry out the resource allocation on (Reference 16).

### 4.4. Other Applications & Systems

4.4.1. <u>Application of P2P Networks</u>

For modeling and analysis of P2P networks such as bitTorrent, the Colored Petri Net are used and as it already mentioned a variety of analysis can be carried out to recognize and simulate the action of the network (Reference 17).

The dependence of the jobs together can be modeled in a distributed P2P system using Petri Net models. Then, these models can be used to carry out the job allocation. In this regard, Petri Net is used for order management in a distribution system (Reference 18).

4.4.2. <u>Application of Multi-Threading in the Operating Systems</u> [5]

"Operating system as the manager of the resources is responsible for allocating the resources to the programs. The number of the requested resources cannot be more than the available resources.

If every process waits for a resource that is possessed by another process, then, there will be a deadlock. Our objective in modeling resource management is to find and resolve the deadlock status. Thus, we should model the requirements so that they show their role in the deadlock. It should be considered that modeling the requests aside from the models of the accessible resources. Therefore, the role of requests must be clarified in the available systems.

The proposed model comprises two sections. The first section shows the process requests and the second section shows the available resources. In this modeling for each process from a separate Petri Net including a transition and two places are used. A place to the possible resource and another place is concerned with the requested resources".[6]

4.4.3. <u>Application in the Service Composition Is Distributed in the Service-Oriented Systems</u>

The Internet was gradually transformed into a place for presenting commercial, organizational, and even Business to Business (B2B) services from a mere trunk. Meaning that the Internet itself is a platform and environment of a distributed computational system from the viewpoint of service-orientation. The discussion of service composition on the Internet provided the ground for modeling this service composition. Service composition can be considered as the equivalent of a workflow. Therefore, Petri Nets can be used to model the service composition and assess the particulars of the efficiency of the resulted system. An article specifically stated that one of these items is examining the existence of deadlock in the system resulted from service composition. Meaning that resource allocation is important in the field of service composition and by Petri Net modeling the status of the system resulted from service composition can be assessed with respect to deadlock and appropriateness of resource allocation. (One of the incentives for using Petri Net in this field is that this model supports composition). Meaning that if we have two separate services of Petri Net models, we can compose these two petri net models and the model pertinent to the composition service will be the composition of these two services). (Reference 19).

---

[5] This subsection which is pertinent to the operating systems is based on reference 25.

[6] Also from Reference 25.

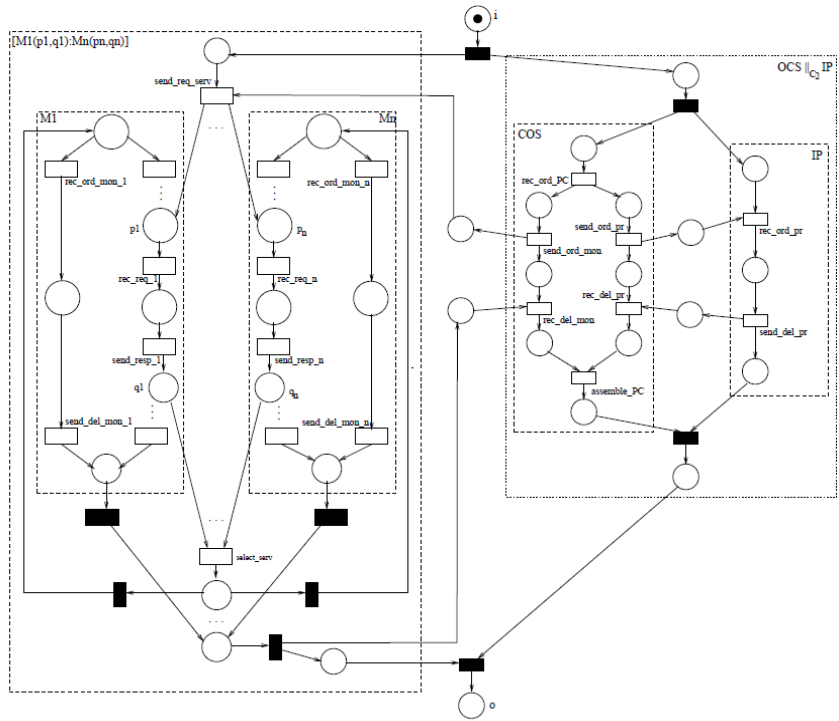

Service $[M_1(p_1, q_1) : M_n(p_n, q_n)] \|_{c_3} (OCS \|_{c_2} IP)$

**Figure 9 - (Reference 19)**

4.4.4. <u>Application in Fault Tolerance and Self-Adjustability of the Distributed Systems</u>
What is the solution when a problem occurs (such as the occurrence of a deadlock)? Here the Petri Net models can be used as well. The Stochastic Petri Nets can be used in Decision Making to achieve Fault Tolerance (Reference 20).

In the distributed pervasive systems, the self-adjustability of the services can be quite important. Especially considering that resources are limited in such systems. Taking into account the resource limitation a variety of self-adjustability scenarios can be proposed. To choose between these scenarios and assess them, the system is modeled using Colored Petri Net models. Then, the system is simulated based on this model, the results of the simulation will be the basis for comparing various scenarios. Using Petri Net, we can select the best scenario from different views, for instance from the self-adjustability view on the resource allocation problem. (Reference 21).

## 5. Conclusion of the Studied Applications

### 5.1. Resource Management Systems

As observed in the earlier chapters, in the Grid environment, workflow systems, multi-agent systems, and operating systems, the resource management subsystems and systems are used and various types of Petri Nets can be used for solving the resource allocation problem.

## 5.2. Predicting & Preventing Deadlock

Both in the introduction chapter and in the subsequent chapters it was mentioned that by relying on the structural concepts in the Petri Net models (such as siphon concept, etc.) The occurrence and possibility of deadlock can be predicted based on the structure of the Petri Net. Predicting deadlock is one of the crucial concerns in the distributed systems and systems based on concurrency.

## 5.3. Automatic Extraction of the Protocol Entities of the Nodes in the Computational Distributed Systems

The Petri Net models are among the modeling and protocol analysis methods. Protocols have a key role in computational distributed systems (Reference 22).

In general, the services provided by a distributed system, are the result of the interaction of a series of protocol entities. Thus, we can make sure that the process of interval sequence of the events develops that the total workflow of the services is executed properly: The required orders are observed together with the concurrencies, etc. Each node in the network of the distributed system possesses its specific protocol entity. It is desirable to automatically extract all descriptions of the protocol entities pertinent to all topics based on the service descriptions, i.e workflow services. Meaning that determined what every node has to do and how to act, that in general and along with concurrency, there will be no deadlock, the system state develops, and finally, submit the requested service to the applicant as a distributed service.

To do so, first, the description of the service is modeled in form of a specific Petri Net called Petri Net plus resisters (PNR). The protocol entities are a PNR, and in fact, they are a Petri Net. Easily and based on the algorithm described in the article, every protocol entity can be extracted from the initial PNR, which is pertinent to the service description. The property of composability of Petri Net is used here as well and PNRs of the entities are determined that the PNR will be their produced composition.

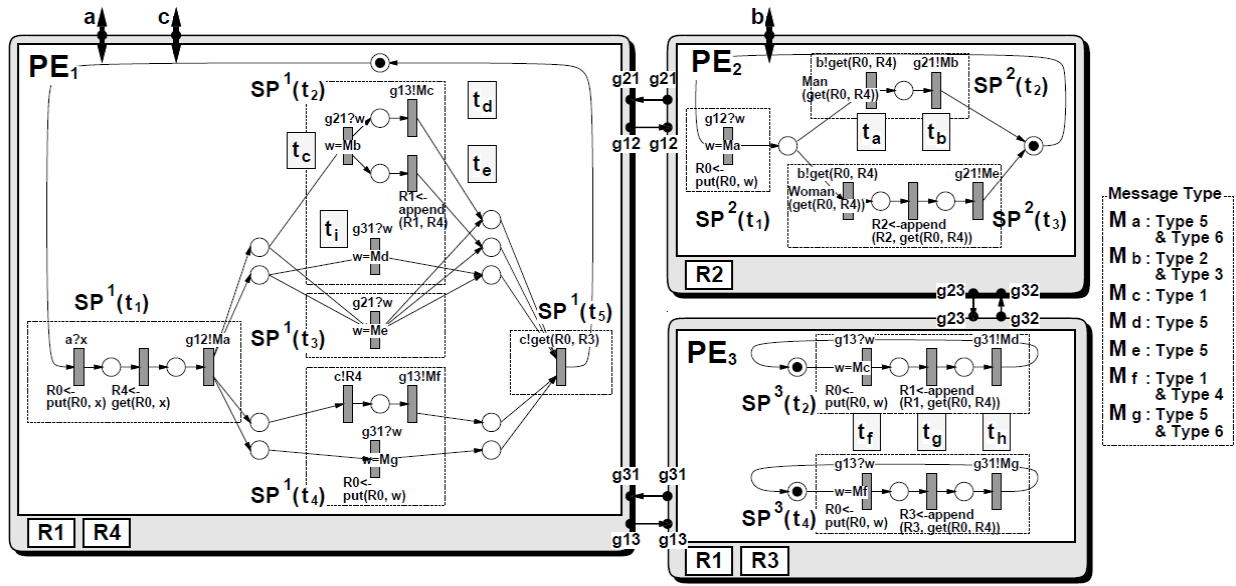

PE₁, PE₂ and PE₃.

**Figure 10 - (Reference 23)**

Here, by the execution of the operations on the structure of Petri Nets, the problem of job distribution on the processing resources are resolved that the concurrency will be without any deadlock (Reference 23).

### 5.4. Assessment of the Efficiency of the System in Resource Allocation and Its Effect on the Total Efficiency of the System

In a distributed computational environment, there is an agreement between energy and efficiency. By accepting the reduction of efficiency, we can use less energy. Sometimes the reduction of efficiency is acceptable. For instance, at night or on the weekend the number of the submitted requests to a mail-server decrease. Therefore, during these times by the tendency to reduce system efficiency against using less energy. This work process (sometimes computes in full capacity and sometimes reduces the capacity of the computational capacity) can be modeled by Stochastic Petri Net. Then, to assess the system efficiency in the adjusted scenario based on it. We are allocating energy resources and Petri Net is used to find the best allocating resource (i.e. The most compatible energy resource allocation proportionate to your workflow). (Reference 24).

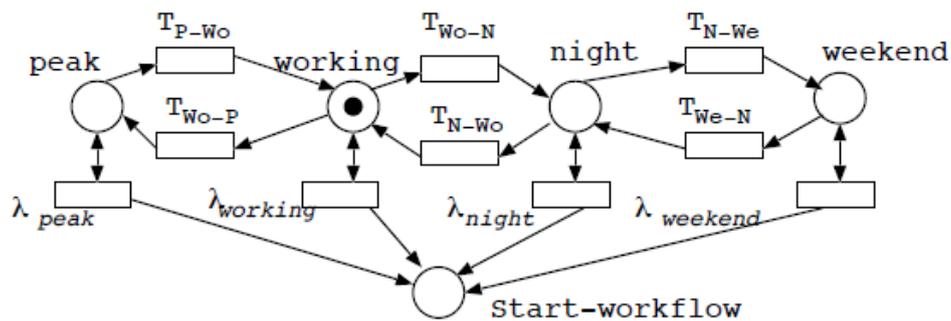

**Figure 11 - (Reference 24)**

## 5.5. Fault Tolerance & Resource Allocation

As seen in the previous chapter, to achieve the fault tolerance, the Petri Net models can be used: Either for modeling the tolerance and adjustability, or prediction and detection of fault (especially to predict deadlock, which is an important system fault), for proper adjustment of the resources of a system through assessment of the efficiency of a resource allocation scenario and receiving feedback in that regard for the future adjustments or for correct allocation of the replication resources in the systems that seek to tolerate fault through replication resources.

## 6. Future Directions

Resource allocation systems are usually modeled as a formal network of flow. There is also a chance for modeling these systems with ccordination semantics (rather than flow semantics). By using Reo Coordination Language, it is possible to gain some new achievments in this application domain of formality and modeling. Reo could be served as a theoretical and also practical basis for protocol design, programming and simulation.